\documentclass[aps,
showpacs,superscriptaddress,citeautoscript,preprint,
prl,floatfix]{revtex4-1}

\usepackage[utf8]{inputenc}

\usepackage{bm}
\usepackage{graphicx}
\usepackage{amsmath,amssymb}
\usepackage{amsfonts}
\usepackage{units}
\usepackage{color}
\usepackage[colorlinks=true,%
            linkcolor=blue,%
            urlcolor=blue,%
            citecolor=blue,%
            filecolor=blue,%
            bookmarksopen=true,%
            pdfauthor={DZ_JPRA_PAO},%
            pdftitle={MBS_BIC_MQDS},%
            pdfsubject={MBS_BIC_MQDS_Manuscript},%
            pdfpagemode=UseOutlines]{hyperref}

\begin{document}
        
\title[]
{
Manipulating photonic signals by a multipurpose quantum junction
}

\author{M.~Ahumada}
\affiliation{Departamento de F\'{\i}sica, Universidad T\'{e}cnica Federico Santa Mar\'{\i}a, Casilla 110 V, Valpara\'{\i}so, Chile}

\author{P.~A.~Orellana}
\affiliation{Departamento de F\'{\i}sica, Universidad T\'{e}cnica Federico Santa Mar\'{\i}a, Casilla 110 V, Valpara\'{\i}so, Chile}

\author{A.~V.~Malyshev}
\affiliation{GISC, Departamento de F\'{\i}sica de Materiales, Universidad Complutense, E-28040 Madrid, Spain}
\affiliation{Ioffe Physical-Technical Institute, 26 Politechnicheskaya str., 194021 St. Petersburg, Russia}


\begin{abstract}
We show that photonic wave packets can be controlled and manipulated in various ways by a multi-functional quantum junction comprising a set of three-level atomic nodes which couple two waveguides. We consider nodes with the $\Lambda$-scheme of the allowed optical transitions, one of which is driven by an external classical electromagnetic control field. Addressing the dynamics of wave packets in such a system, we demonstrate that an optical pulse can be routed into one of the selected output channels, split into several parts, delayed by a desired time or stored partially in the junction. These modes of operation can be selected and controlled by the external classical field. We argue, therefore, that our proposed design is a promising prototype of a multipurpose quantum junction.

\end{abstract}

\maketitle

The operation of modern networks relies on the functionality of rather complex key devices, such as switches, routers, and repeaters or amplifiers, etc. The functionality of these devices is provided by combinations of elementary operations performed by much more basic elements, such as delay lines, memory cells, and simple routing nodes. Although this technology is well developed for traditional classical networks, it is still an active research area in the case of quantum networks, which leaves some room for improvement.

In a quantum network, photons are believed to be the natural candidates for carrying information with high fidelity as flying qubits in long-distance communications over quantum channels \cite{kimble2008quantum,Northup2014,Monroe2002,ladd2010quantum,Ritter2012,flamini2018photonic,OBrien2009}. Therefore, considerable effort has been put into studies of photon transport in one-dimensional waveguides coupled to quantum emitters \cite{RevModPhys.87.1379,PhysRevLett.101.100501,PhysRevA.78.053806,Chen_2011}. These emitters can be coupled to the quantum channels not only to inject information carriers but also to manipulate and route them, controlling pathways of signals in the network. Routing is one of the most important operations in a network, for which reason various implementations of quantum routers has been proposed~\cite{PhysRevLett.111.103604,PhysRevA.89.013805,Huang2018,J.Huang,Lu:15,Liu2016,Aoki2009,PhysRevX.3.031013,Shomroni903,Li2016,Cao:17,Yan2014,PhysRevA.97.023821,PhysRevLett.107.073601,Yuan2015,Hu2017,nanolett8}.

It would also be advantageous if a single device could provide all fundamental functionality, operating as a router, a delay line, a splitter, or an information storage node. Below we report on a prototype of such a multipurpose device: a quantum junction comprising three-level atoms coupling two waveguides. One of the allowed optical atomic transitions is driven by an external classical electromagnetic control field, which can be used to select the operating mode of the device and manipulate optical signals propagating in the waveguides. 

Our proposed prototype device comprises two waveguides, which are symmetrically coupled by a set of $N_a$ sequential three-level systems. The $\mathrm A$ and $\mathrm B$ waveguides are modeled as one-dimensional arrays of sites described by the bosonic operators $a_n^\dagger$ and $b_n^\dagger$, which create a photon with the energy $\hbar\omega_0$ at the $n$-th site of the corresponding waveguide. The sites are coupled by the constant nearest-neighbor interaction $\xi$. For simplicity, we consider identical waveguides and three-level systems. We assume also that the latter have the $\Lambda$ type-level scheme, comprising the state $|g_j\rangle$, the excited $|e_j\rangle$, and the third state $|s_j\rangle$ (where $j$ labels the atom, $j=1\ldots N_a$). The energies of these atomic states are $E_g$, $E_e$, and $E_s$, respectively. The dipole-allowed transitions $|g_j\rangle\leftrightarrow|e_j\rangle$ are coupled to the modes $a_j^\dagger$ and $b_j^\dagger$ of the neighbouring waveguide sites with the coupling constant $g$. Other allowed transitions $|s_j\rangle\leftrightarrow|e_j\rangle$ are driven by the external classical control field with the frequency $\omega_c$ and the Rabi frequency $\Omega$. Transitions $|g_j\rangle\leftrightarrow|s_j\rangle$ are dipole forbidden.

Within the rotating frame approximation and in the rotating frame with respect to
\begin{eqnarray}
H_R&=&\hbar\omega_0\sum_n
(\hat{a}^{\dag}_{n}\hat{a}_{n}+\hat{b}^{\dag}_{n}\hat{b}_{n})
\nonumber 
\\ \nonumber
&+&\hbar\omega_0\sum_j
(\vert e_{j}\rangle\langle e_{j}\vert+\vert s_{j}\rangle\langle s_{j}\vert)
-\hbar\,\omega_c\sum_j
|s_j\rangle\langle s_j|\ ,
\end{eqnarray}
the Hamiltonian of the system reads as
\begin{eqnarray}\label{H}
H&=&
-\xi\sum_n
\Big[
\hat{a}^{\dag}_{n}\hat{a}_{n+1}+
\hat{b}^{\dag}_{n}\hat{b}_{n+1}+\mathrm{H.c.}\Big]
\nonumber
\\
&+&\sum_j
\Big[\Delta_{e}\vert e_{j}\rangle\langle e_{j}\vert+(\Delta_{e}-\Delta_{s})\vert s_{j}\rangle\langle s_{j}\vert\Big]
\\
\nonumber
&+&\sum_j
\Big[g\,\vert e_{j}\rangle\langle g_j\vert(\hat{a}_{j}+\hat{b}_{j})+\hbar\Omega\vert e_{j}\rangle\langle s_{j}\vert+\mathrm{H.c.}\Big]  
\ ,
\end{eqnarray}
where $\Delta_{e}=E_e-E_g-\hbar\omega_0$ and $\Delta_{s}=E_e-E_s-\hbar\omega_c$ are the detunings of the photon energies from the two allowed transition energies \footnote{Note that this Hamiltonian is different from those in some earlier works due to differences in the used rotating frames.}. Hereafter, we consider the resonant case $\Delta_e=\Delta_s=0$, we also set $\hbar\omega_0$ as the reference energy level, the coupling constant $\xi$ as the energy unit, and $\hbar=1$. Each standalone waveguide supports plane wave modes with the dispersion relation $E(k)=-2\,\cos{k}$, where the dimensionless wave vector $k\in[0,\pi]$; the center of the band, $E(\pi/2)=0$, corresponds to the photon energy $\hbar\omega_0$. 

To study the dynamics of a wave packet defined as $\vert \Psi \rangle=\sum_n\big[\alpha_n a_n^\dagger|g_n, 0\rangle + \beta_n b_n^\dagger|g_n, 0\rangle\big] + \sum_j\big[u_j|e_j, 0\rangle+v_j|s_j, 0\rangle\big]$,
with $|0\rangle$ being the vacuum state of the waveguides, we use the time-dependent Schr\"odinger equation with the Hamiltonian (\ref{H}) written for the amplitudes $\alpha_n$, $\beta_n$, $u_j$, and $v_j$:
\begin{eqnarray}\label{sys}
\dot\alpha_n&=&i\,(\alpha_{n+1}+\alpha_{n-1})-i\,g\,\delta_{nj}\,u_j \ ,
\nonumber\\
\dot\beta_n&=&i\,(\beta_{n+1}+\beta_{n-1})-i\,g\,\delta_{nj}\,u_j \ ,
\\ \nonumber
\dot u_j&=&-i\,\Omega\,v_j-i\,g\,(\alpha_j+\beta_j) \ ,
\\ \nonumber
\dot v_j&=&-i\,\Omega\,u_j\ ,
\end{eqnarray}
where $\delta_{nj}$ is the Kronecker symbol, $j=1\ldots N_a$, $n=-N\ldots N+N_a$, and $N$ is the number of sites in each of the left and right channels (branches) of the waveguides. For further reference, we define different regions of the system as follows: the left channels $\mathrm{A_{L}}$ and $\mathrm{B_{L}}$ (with $n=-N+1\ldots 0$) of the waveguides $\mathrm{A}$ and $\mathrm{B}$ respectively, the central or connected region $\mathrm{C}$ (with $n,j=1\ldots N_a$), and the right channels $\mathrm{A_{R}}$ and $\mathrm{B_{R}}$ (with $n=N_a+1\ldots N+N_a$). We also define integrated densities of probability to find the wave packet in different waveguide channels and at the atomic states: 
\begin{equation}
\label{probs}
P_{AL,AR}=\sum_{n\in \mathrm{A_{L,R}}}\vert \alpha_n\vert ^2\ , \quad P_{BL,BR}=\sum_{n\in \mathrm{B_{L,R}}}\vert \beta_n\vert ^2\ , \quad
P_{C}=\sum_{j}\vert u_j\vert ^2+\vert v_j\vert ^2\ .
\end{equation}

We solve the system (\ref{sys}) numerically using the following normalised Gaussian wave packet as the initial condition:
\begin{equation}
\label{psi0} 
\alpha_n=\frac{1}{\sqrt{\sigma\sqrt{\pi}}}\;e^{-{\frac{(n-n_{0})}{2\sigma^{2}}^{2}}+i\,k_{0}n},
\quad \beta_n=u_j=v_j=0\ ,
\end{equation}
where $\sigma$, $k_0>0$, and $n_0$ are the width, the wave vector, and the initial position of the center of the wave packet, respectively. Such a wave packet is propagating from left to right in the  $\mathrm{A_L}$ channel. We always choose $n_0$ in such a way that the amplitude of the wave packet at the connected region $\mathrm C$ is negligible at $t=0$, typically $n_0=-[N-3\sigma]$. 

Hereafter, we consider the system with $N=1000$, $N_a=12$, and $g=0.5$. Such a system with $12$ atoms forming the junction was studied recently in Ref.~\cite{Ahumada2019}, where it is demonstrated that it has promising scattering properties in the stationary regime for certain sets of parameters. In this paper, we address the dynamics of wave packets for the most relevant parameter sets.

\begin{figure*}[t]
\includegraphics[width=\textwidth]{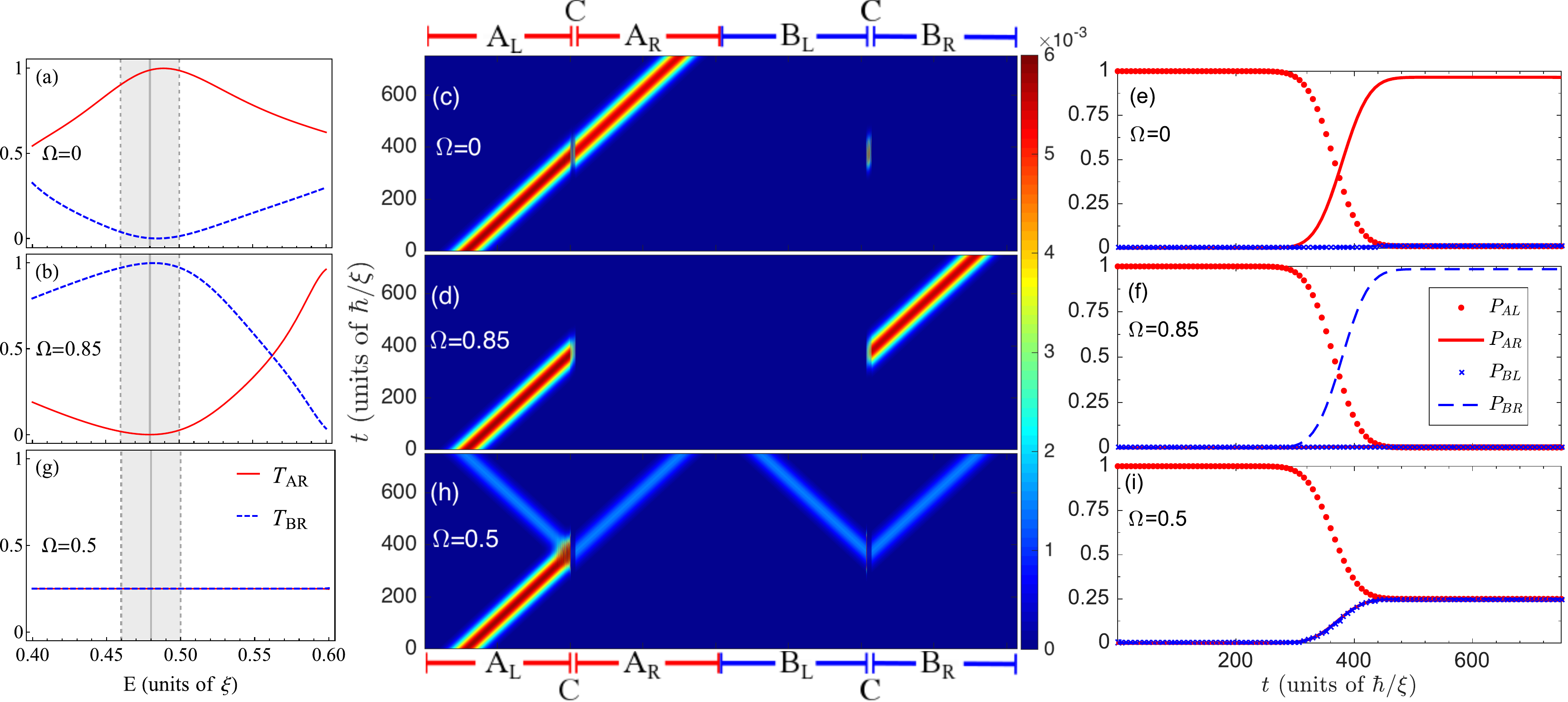}
\caption{Stationary characteristics and corresponding wave packet dynamics calculated for three values of the control field: $\Omega=0$ (upper row), $\Omega=0.85$ (middle row), and $\Omega=0.5$ (lower row). Left column: stationary spectra of the transmission $T_{AR}$ into the $\mathrm{A_R}$ channel (solid red line) and $T_{BR}$ into the $\mathrm{B_R}$ channel (dashed blue line). The vertical grey line indicates the energy $E_{0}=0.48$ of the center of the wave packet whose other parameters are $k_{0}=\arccos{(-E_{0}/2)}\approx 1.33$ and $\sigma=100$. The shaded area indicates the width of the wave packet in the energy space $\Delta \approx 0.02$. Middle column: the spatio-temporal map of the wave packet probability density $|\Psi|^{2}$, where the waveguide regions are ordered in the following sequence: $\mathrm{A_L, C, A_R, B_L, C, B_R}$. Right column: the integrated probability densities $P_{AL}$ (dotted red line), $P_{AR}$ (solid red line), $P_{BL}$ (crossed blue line), and $P_{BR}$ (dashed blue line).}
\label{FigRoutingSplitting}
\end{figure*}

First, we investigate the system in the regime of controlled routing. We show that a wave packet incoming from the left input channel $\mathrm{A_L}$ can be routed into one of the two right output channels, $\mathrm{A_R}$ or $\mathrm{B_R}$, which can be selected by the control field $\Omega$. To this end, we start with inspecting the stationary spectra of the transmission into the two right output channels, $T_{AR}$ and $T_{BR}$, for two different values of the control field: $\Omega=0$ and $\Omega=0.85$ (see Ref.~\cite{Ahumada2019} for details of their calculation); the spectra are shown in panels (a) and (b) of Fig.~\ref{FigRoutingSplitting}, respectively. These spectra have useful features in the vicinity of $E_0\approx 0.48$, i. e., the high and low values of the transmission probabilities into the output channel $\mathrm{A_R}$ or $\mathrm{B_R}$ can be swapped when the control field is switched from $\Omega=0$ to $\Omega=0.85$ or vice versa. Therefore, if the incoming wave packet is properly centered at $E_0\approx 0.48$ and if its width in the energy space $\Delta E$ is smaller than the width of the transmission features ($\approx 0.1$), then almost the whole wave packet can be routed into one or the other output channel, which is selected by the control field $\Omega$. 

To demonstrate the feasibility of such routing, we consider the dynamics of the wave packet (\ref{psi0}) centered at $k_0=\arccos{(-E_0/2)}\approx 1.33$ and having the width $\Delta E=0.02$ in the energy space (which is much smaller than the width of the above mentioned spectral features). The corresponding wave packet width in the real space $\sigma$ can be estimated as $\sigma=2\,\sin{k_0}/\Delta E\approx 100$. The dynamics of such a wave packet for $\Omega=0$ and $\Omega=0.85$ is presented in panels (c) and (d) of Fig.~\ref{FigRoutingSplitting}, respectively. These panels show the spatio-temporal maps of the wave packet probability density $|\Psi|^{2}$, manifesting apparently efficient routing. Panels (e) and (f) of Fig.~\ref{FigRoutingSplitting} demonstrate the efficiency of the routing at a more quantitative level: they show the dynamics of the integrated probabilities $P_{AR}$ and $P_{BR}$, whose asymptotic values are swapped between approximately $0.985$ and $0.015$ when the classical field switches between $\Omega=0$ and $\Omega=0.85$. Thus, the incident wave packet can be routed into one or the other output channel selected by the control field $\Omega$.

Second, we address the system in the wave packet ``1/4-splitting'' regime. As was demonstrated recently~\cite{Ahumada2019}, the stationary transmission and reflection spectra of the considered system have wide flat sub-bands in the vicinity of $E=\pm\Omega$. Within these sub-bands, the probabilities of scattering into the four possible channels ($\mathrm{A_{L}}$, $\mathrm{A_R}$, $\mathrm{B_L}$, and $\mathrm{B_R}$) are all approximately equal to $1/4$ (see panel (g) of Fig.~\ref{FigRoutingSplitting} where these characteristics are shown for $\Omega=0.5$). Thus, if a propagating wave packet is properly centered at $E\approx\pm\Omega$ and its energy width is less than that of the flat sub-band, which is on the order of $2g=1$ in our case, the packet will be split into four approximately equal parts. The dynamics of such a wave packet was calculated for $\Omega=0.5$ with the same parameters as above. The results are presented in panels (h) and (i) of Fig.~\ref{FigRoutingSplitting}, which show an even splitting of the incident wave packet. In particular, the right panel (i) shows that the asymptotic values of all integrated probability densities are equal: $P_{AL}=P_{AR}=P_{BL}=P_{BR}=0.25$, indicating that the incident wave packet is split into four equal parts.

\begin{figure}[t]
\includegraphics[width=10cm]{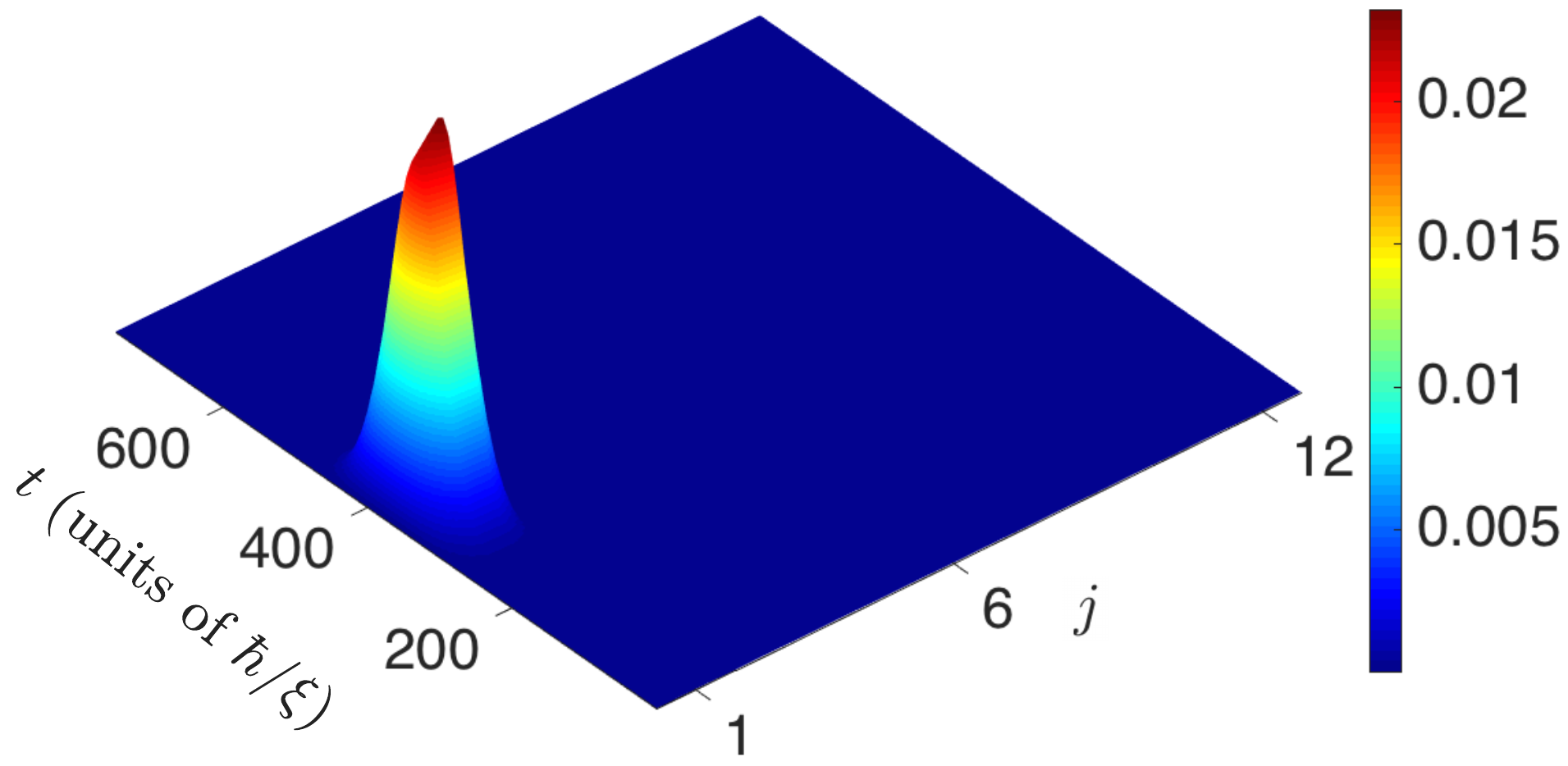}
\caption{The dynamics of the probability densities at the atomic states $|e_j\rangle$ and $|s_j\rangle$ organized in couples $\{\vert u_j \vert ^2,\,\vert v_j\vert ^2\}$ ($j=1\ldots 12$), demonstrating the point-like excitation of the extended junction (see text for details). All parameters are the same as for the lower row of Fig.~\ref{FigRoutingSplitting}.} 
\label{FigScattering}
\end{figure}

To get an insight in the mechanism of such an even splitting we plot in Fig.~\ref{FigScattering} the map of the probability densities $|u_j|^2$ and $|v_j|^2$ at the atomic states $|e_j\rangle$ and $|s_j\rangle$. The map shows clearly that only the states of the leftmost atom (with $j=1$) are excited during the pulse scattering. The latter suggests that the whole junction acts as a point scattering defect, and therefore the scattering is symmetric in this case: the amplitude of such a scattering into all the four channels is the same. This finding confirms our earlier result~\cite{Ahumada2019} obtained for the stationary case: 
within the $1/4$-scattering flat bands, the whole junction acts as a very high quantum mechanical barrier;  then, the amplitude of the wavefunction at the atomic states decays exponentially, and it is already negligible at the second atom ($j=2$). Thus, the whole extended junction is acting as if it was a point defect, splitting the incoming pulse into four approximately equal parts. The latter suggests a mechanism of an even and efficient optical pulse splitting. On the other hand, in the single-photon regime, such a system can be used as a quantum random number generator~\cite{RevModPhys.89.015004,Grafe2014}: each mutually exclusive single-photon detection in either of the four channels would give a two-bit random number.

\begin{figure*}[t]
\includegraphics[width=\textwidth]{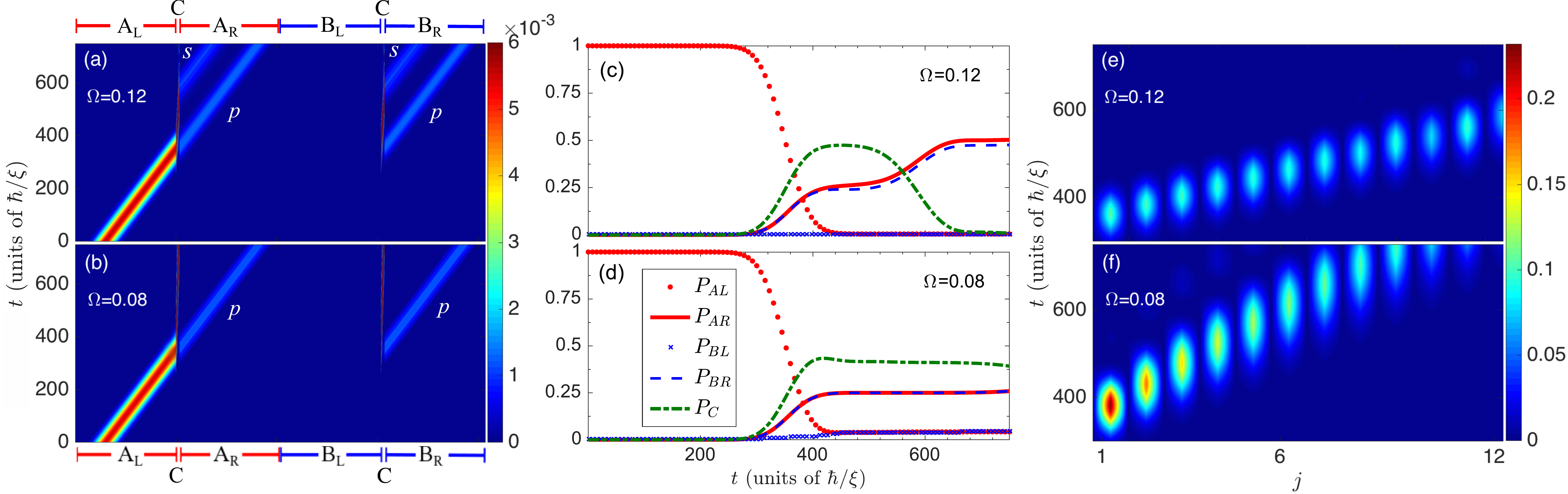}
\caption{Probability density dynamics calculated for two different values of the control field: $\Omega=0.12$ (upper row) and $\Omega=0.08$ (lower row). Left column: the spatio-temporal map of the wave packet probability density $|\Psi|^{2}$, where the waveguide regions are ordered in the following sequence: $\mathrm{A_L, C, A_R, B_L, C, B_R}$. The {\it primary} and {\it secondary} pulses are labeled by $p$ and $s$, respectively. The middle column shows the integrated probability densities $P_{AL}$ (dotted red line), $P_{AR}$ (solid red line), $P_{BL}$ (crossed blue line), $P_{BR}$ (dashed blue line), and $P_{C}$ (green dash-dotted line). Right column: the dynamics of the probability densities at the states $|e_j\rangle$ and $|s_j\rangle$ organized in couples $\{\vert u_j \vert ^2,\,\vert v_j\vert ^2\}$ ($j=1\ldots 12$), demonstrating the propagation of the {\it secondary} pulse over the atomic states (see text for details). The wave packet parameters are $k_{0}=\pi/2$ and $\sigma=100$.}
\label{FigStorage}
\end{figure*}

Finally, we address the delay and storage properties of the quantum junction. To demonstrate these, we simulate the propagation of the wave packet (\ref{psi0}) with $\sigma=100$ and $k_0=\pi/2$ (such a wave packet is centered at the center of the waveguide energy band, {\it i. e.} at $\hbar\omega_0$). Figure~\ref{FigStorage} presents results of such calculations for $\Omega=0.12$ in the upper row and those for $\Omega=0.08$ in the lower row. The left column of Fig.~\ref{FigStorage} demonstrates the spatio-temporal map of the wave packet probability density $|\Psi|^{2}$. To describe the dynamics more quantitatively, we also plot the integrated probability densities $P_{AL,AR}$, $P_{BL,BR}$, and $P_{C}$ in the middle column of Fig.~\ref{FigStorage}.

The left and middle columns of Fig.~\ref{FigStorage} show that when the wave packet (incoming from the $\mathrm{A_L}$ channel) reaches the junction, it is partially scattered into the two right output channels $\mathrm{A_R}$ and $\mathrm{B_R}$. As a result, two almost identical {\it primary} scattered pulses start propagating freely in the output channels [these are labeled by $p$ in Figs.~\ref{FigStorage}(a) and \ref{FigStorage}(b)]. Each of these two pulses carries about $1/4$ of the integrated probability density; the rest $1/2$ of the density is stored in the atomic states as can be seen from panel (c) of Fig.~\ref{FigStorage} for $400 \lesssim t \lesssim 500$ (see the green dash-dotted line). Panels (a) and (c) of Fig.~\ref{FigStorage} show also that two {\it secondary} pulses start propagating in the two right output channels after some delay [those are labeled by $s$ in Fig.~\ref{FigStorage}(a)]. These {\it secondary} pulses are broadened with respect to the {\it primary} ones but they also carry about $1/4$ of the integrated probability density each [see Fig.~\ref{FigStorage}(c)]. 

To get an insight in the mechanisms of the {\it secondary} pulse formation and delay we plot in the right column of Fig.~\ref{FigStorage} the spatio-temporal maps of the probability densities $|u_j|^2$ and $|v_j|^2$ at the atomic states $|e_j\rangle$ and  $|s_j\rangle$, respectively. The two maps show that the incident pulse is forming the {\it secondary} one, which is propagating in the junction over the atomic $|s_j\rangle$ states (note that $|u_j|^2=0$). While such {\it secondary} pulse is propagating in the junction, it has a considerably smaller group velocity than those pulses propagating freely in the input/output channels. The group velocity is small because the $|s_j\rangle$ states are coupled only indirectly, via the waveguide and the excited $|e_j\rangle$ states. The $|s_j\rangle$ and $|e_j\rangle$ states are coupled by the field $\Omega$ and, therefore, it is natural to expect that the smaller group velocity depends on the control field $\Omega$. Panels (e) and (f) of Fig.~\ref{FigStorage} confirm this expectation, showing that the group velocity decreases with the coupling field (note the difference in the slopes in panels). When the slow {\it secondary} pulse reaches the right extreme of the junction it scatters into the two right output channels, forming two almost identical broadened {\it secondary} scattered pulses which propagate freely (with the normal group velocity) in $\mathrm{A_R}$ and $\mathrm{B_R}$ channels following the {\it primary} pulses. Because of the aforementioned difference in the group velocities, the {\it secondary} pulses are delayed with respect to the {\it primary} ones. As we have argued, this delay can be controlled by the classical field $\Omega$, which suggests a mechanism of the pulse delay control.

Note also that when the incident pulse has already scattered into the output channels in the form of the {\it primary} pulses, the {\it secondary} pulse can still be propagating in the junction over its atomic states $|s_j\rangle$. If the control field $\Omega$ is switched off at such a moment, the {\it secondary} pulse can be ``frozen'' or stored in the atomic states $|s_j\rangle$ because they would be completely decoupled from the states $|e_j\rangle$ and the rest of the system. Within our idealized model (which neglects dissipation completely) such storage has an unlimited time: the stored part of the wave function preserves its amplitudes at the atomic states and relative phases. Then, if the external field is switched back on, the stored pulse would be ``released'' and continue its propagation, which suggests a mechanism of the pulse trapping or storage control. A more detailed study of the pulse trapping or storage and group velocity control is due to be published elsewhere.

In conclusion, we have studied the dynamics of photonic pulses in the system of two waveguides coupled by the multipurpose quantum junction comprising a set of three-level atoms with the $\Lambda$-scheme of the allowed optical transitions, one of which is driven by an external classical electromagnetic control field. We demonstrate that photonic wave packets propagating in the system can be controlled and manipulated in various ways. In particular, an incident pulse can be routed into a selected output channel or split in several parts, some of which can be delayed by an amount of time determined by the control field. The pulse can also be partially trapped or stored in the junction and released afterward in a controlled way. All these operations can be performed with high efficiency by the same physical device. We argue, therefore, that our proposed model system can provide useful guidelines on the design of the multi-functional junctions, making the future all-optical circuitry building blocks more multipurpose and integrated. Given that our model is simple and quite generic, similar devices can be designed based not only on atomic but also on other physical three-level systems, such as SQUIDs. We believe that the wide variety of operational regimes combined with high efficiency of operation makes our model design a promising prototype proposal for applications in next-generation information processing and communication technologies.


Work in Madrid was supported by MINECO grant MAT2016-75955. M.~A. and P.A. acknowledge financial support from DGIIP UTFSM and FONDECYT Grant No. 1180914.  M.~A. also acknowledges support from CONICYT Doctorado Nacional through Grant No. 21141185 and J.~F. Mar\'in for fruitful discussions. A.~M. is grateful to V. A. Malyshev for critical discussions.

\bibliographystyle{apsrev4-1}
\bibliography{references}

\end{document}